\documentstyle[psfig,times]{mn}
\begin{document}
\hsize=6truein

\renewcommand{\thefootnote}{\fnsymbol{footnote}}
\newcommand{\beppo}{{\it BeppoSAX} }
\newcommand{\rosat}{{\it ROSAT} }

\title[\beppo --  \rosat PSPC observations of A3562]
{\beppo -- \rosat PSPC observations of the 
Shapley Supercluster: A3562}

\author[S. Ettori et al.]
{\parbox[]{6.in} {S. Ettori$^1$, S. Bardelli$^2$, S. De Grandi$^3$, 
S. Molendi$^4$, G.~Zamorani$^2$, E.~Zucca$^2$ \\
\footnotesize
$^1$ Institute of Astronomy, Madingley Road, CB3 0HA Cambridge, England \\
$^2$ Osservatorio Astronomico di Bologna, via Ranzani 1, I-40127 Bologna, 
Italy \\
$^3$ Osservatorio Astronomico di Brera, Via Bianchi 46, I-23807 Merate
(LC), Italy \\
$^4$ Istituto di Fisica Cosmica ``G.Occhialini'', Via Bassini 15, I-20133
Milano, Italy \\
}}                                            
\maketitle

\begin{abstract}
We present a combined analysis of the \beppo and \rosat PSPC
observation of the cluster of galaxies A3562, a massive 
member of the core of the Shapley Supercluster.
With a complex and interacting structure composed from two groups
of galaxies and A3558 to the West, the surface brightness 
of A3562 shows excess in the sectors to East and South when
compared with an azimuthally averaged model of the emission.
The emission tends to be flatter, and the distribution of the gas 
broader, along the merging axis and in opposition to the two groups.
We present the first determination of the gradients of the gas temperature 
and metallicity for a cluster in the Shapley region at large 
distance from cluster center.
From an analysis of the \beppo data in annuli and sectors, we observe
both the profiles to be flat within 8\arcmin\ ($\sim$ 0.62 Mpc), with
an emission-weighted values of $kT = 5.1 \pm 0.2$ keV and 
$Z = 0.39 \pm 0.05 Z_{\odot}$. The value of the temperature is 
consistent with recent {\it ASCA} measurements and is significantly
higher than previous estimates obtained from \rosat and {\it EXOSAT}.
We discuss the possible reasons of this disagreement.
Between 8\arcmin\ and 20\arcmin, the plasma temperature declines 
to about 3.2 keV. When a polytropic profile is used to represent
the gas temperature profile, the best fit polytropic index 
is $1.16 \pm 0.03$.
These results imply a total mass within the virial radius of 
$3.9 (\pm 0.4) \times 10^{14} h_{50}^{-1} M_{\odot}$, 
between 40 and 80 per cent lower than the optical estimate, 
and a gas mass fraction of about 30 per cent. 
\end{abstract}


\section{INTRODUCTION} 
Superclusters of galaxies are the largest and more massive
structures in the Universe and have a dynamical status
that provides direct information on the evolution of 
structures formed by accretion of matter in a hierarchical 
bottom up scenario.
Numerical simulations on scales of cosmological relevance (e.g. 
Cen \& Ostriker 1994) reveal that merging in a Cold Dark Matter 
dominated Universe happen along preferential directions at whose 
intersection rich clusters and core of Superclusters form.

The Shapley Supercluster is a remarkable concentration of galaxies
and clusters of galaxies in the nearby Universe 
(Raychaudhury 1989, Scaramella et al. 1989, 
Raychaudhury et al. 1991, Zucca et al. 1993) 
and presents a highly overdense core formed by three ACO (Abell et al. 
1989) clusters, A3556, A3558 and A3562. This structure
is elongated by $\sim 3^{\circ}$ in the East--West
direction with a comoving size of about 15 $h_{50}^{-1}$ Mpc.
The size of the core region is typical for a structure that is collapsing,
with A3556 and A3562 that are on the way to merge with the more
massive A3558 (Metcalfe et al. 1994, Bardelli et al. 1994, 
Ettori et al. 1997, Hanami et al. 1999).
Considering how the region is undergoing strong interaction among 
the component clusters, it represents an ideal place in the local 
Universe to investigate the dynamical status of massive structures.

For these reasons, several  multiwavelength studies have been 
dedicated to the Shapley Concentration of galaxies since the very
first observation by Shapley in 1930.
In the recent years,  Bardelli et al. (1994, 1998a, 1998b, 2000), 
Quintana et al. (1995, 1997) and Drinkwater et al. (1999) have 
investigated its optical properties.
Venturi et al. (1997, 1998) have analyzed the radio emission from
galaxies in the region and its interaction with the intracluster gas. 
In the X-ray energy band, analyses of the Shapley supercluster 
have been performed with {\it GINGA} (Day et al. 1991), {\it Einstein}
(Raychaudhury et al. 1991, Breen et al. 1994), pointed (Ettori et al. 1997) and
All-Sky Survey (Kull \& B\"ohringer 1999) \rosat PSPC observations
and {\it ASCA} (Hanami et al. 1999) data to
describe both spectroscopically and spatially its extended emission.

In particular, the study of the distribution of the hot plasma 
in the X-ray waveband within each cluster and its environment can 
provide detailed informations on the merging history of the 
Shapley Supercluster.
A detailed analysis of the most massive cluster, A3558,
was based on spatial and spectral analysis of a \rosat PSPC
observation (Bardelli et al. 1996) and {\it ASCA} spectra
(Markevitch \& Vikhlinin 1997).
From {\it ASCA} data, Hanami et al. (1999) have proposed a scenario 
of the evolution 
sequence of the clusters in the Shapley Supercluster, from an
ongoing merger (SC1329-313), to a post-merger cluster (A3556), 
to the relaxed poor cluster (SC1327-312) and to the rich clusters 
with mergers in succession (A3558 and A3562).
Kull \& B\"ohringer (1999) have found evidences of a physical filamentary
structure of hot gas embedding the three clusters (A3558, A3556, A3562).

The scenario of the merging evolution of the A3558-A3556-A3562 structure,
as the remnant of a cluster-cluster collision seen just after the
first core-core encounter, 
is also supported from an analysis of galaxy colors and substructures
in Bardelli et al. (1998b).

With the present work, we start an investigation of the members
of the core of the Shapley Supercluster combining the spatial
resolution of \rosat PSPC with the higher spectral capabilities 
of the Italian-Dutch satellite \beppo$\!$.

We present in Table~1 the X-ray observations of A3562 discussed
in this work. The \rosat PSPC data were retrieved from the archive,
while the \beppo observation has been our AO2 target.
All the errors quoted are at the 68.3 per cent confidence level
($1 \sigma$) unless otherwise stated.
Hereafter, we assume a cosmological model with a Hubble constant,
$H_0$, of $50 h_{50}^{-1}$ km s$^{-1}$ Mpc$^{-1}$, the density
parameter, $\Omega$, equal to 1 and $\Lambda = 0$.
Therefore, $1'$ corresponds to a proper radius of 78 kpc at the
redshift of A3562 of $0.0483^{+0.0008}_{-0.0010}$ (Bardelli et al.
1998a). 

The paper is organized as follows: in Section~2, we discuss the
spatial morphology of the X-ray emission from A3562 and the results
of the spectral analysis using the \rosat PSPC observation;
in Section~3, the \beppo spectra are analyzed. 
The constraints on the masses obtained from the observed 
gas temperature profile are presented in Section~4.
Finally, we discuss the main conclusions of this work 
in Section~5. 

\begin{table}
\caption[]{Observations summary of A3562.
}
\begin{tabular}{ccccc}
Satellite & Detector & Band & Date & Exposure (ksec)   \\
\rosat  &  PSPC B  &  0.1--2.4 keV & 1993 Jan 19 & 20.2 \\
\beppo  &  LECS  &  0.1--4 keV &  1999 Jan 31 & 18.9 \\
\beppo  &  MECS 2+3  &  2--10 keV &  1999 Jan 31 & 45.9 \\
\end{tabular}
\end{table}

\section{\rosat PSPC data}
The \rosat (the ROentgen SATellit, Tr\"umper 1983) 
Position Sensitive Proportional Counter (PSPC) has a field of view
with a diameter of $2^{\circ}$ and is divided by the ribs into 
(i) an unobscured central circular region of 20 arcmin in radius, 
and (ii) an outer annulus sectioned in eight equal sectors.
The FWHM of the on-axis Point-Spread-Function (PSF) is of about
25 arcsec. The energy resolution of the PSPC is $dE/E = 0.43
(E/0.93)^{-0.5}$ (FWHM) over the entire sensitive area of the detector
and in the energy band 0.1--2.4 keV.

The reduction and analysis of these data have made use of the
{\sc Ftools} version~4.2 package, {\sc Xselect} 1.4b and 
{\sc Xspec} 10.00 (Arnaud 1996).

\subsection{Spatial analysis}

The PSPC images with pixel size of 15\arcsec have been constructed from
counts in the 0.5--2 keV
band, where the Galactic and particle background are minimum, after
correction for instrumental and telemetry dead time, and exclusion of times 
of high background counts. The total livetime is 16,597 seconds.
We also require the Master Veto count rate to be
less than 170 counts s$^{-1}$ (cf. guidelines for reduction of PSPC data
in Snowden et al. 1994). Using the \rosat Interactive Data Language
(IDL)  user-supplied libraries, we have divided the images by the
respective exposure maps in the same energy band and corrected them for
vignetting and exposure-time. The region of the detector support rib has
been masked out as well all detected point sources. 

\begin{figure}
\psfig{figure=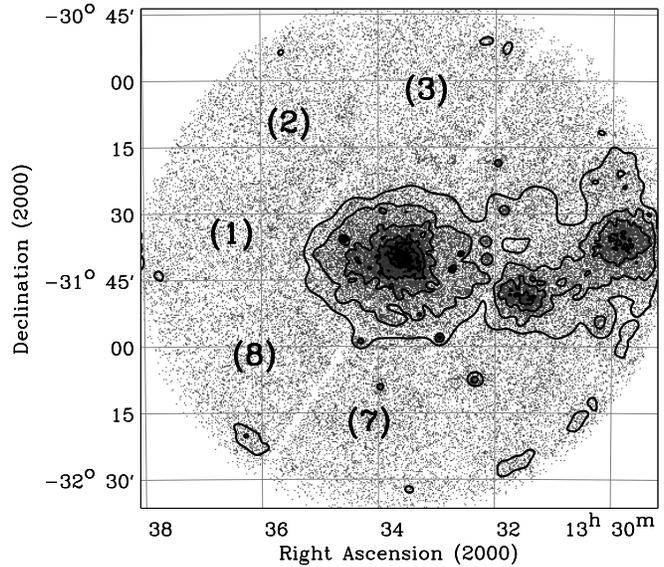,width=.5\textwidth}
\caption[]{ \rosat PSPC raw data of A3562. 
The image has been adaptively smoothed to consider
in the smoothing kernel
structures with a significance larger than 5 times the
local background. Here, we overplot to the raw image the eight contours
of the smoothed image with values equally spaced in logarithmic scale
between the maximum value in the background map of 0.4 cts pix$^{-1}$
and the maximum value of the smoothed image of 67.6 cts pix$^{-1}$.
The labels '(1)', '(2)', ..., '(8)' indicate clockwise the 
sectors used for the spatial analysis. 
To the West, towards the center of the Shapley core, two groups of 
galaxies are evident, SC1329-313 and SC1327-312, respectively.
} \label{fig:asm}
\end{figure}

The raw image has been smoothed in IDL with an adaptive kernel
method, ASMOOTH (Ebeling, White \& Rangarajan 1999),
using a Gaussian kernel with size between 0.2 and 74.8 pixels
and a characteristic smoothing threshold
of 5 times the map of the local background (Fig.~\ref{fig:asm}).
The contoured image shows clearly the interactions in act between
A3562 and the two groups westwards. 

The local background (and the respective error) is estimated from
the average of the counts in the three exposure-corrected 
$45^{\circ}$-sectors located eastward (cf. Sectors~1, 2 and 8 in 
Fig.~\ref{fig:asm}), in opposition to the two groups, 
between 40\arcmin\ and 45\arcmin\ from the X-ray emission center 
(RA, Dec) = 13$^{\rm h}$ 33$^{\rm m}$ 35\fs5, 
--31\degr 40\arcmin 05\arcsec. 
The choice of these Sectors is driven from the necessity to reduce
the contamination from the extended sources to the West that
seems to affect 5 out of 8 sectors in different proportion (see
Table~2). 

We measure an average local background 
of $4.06 (\pm 0.10) \times 10^{-4}$ cts s$^{-1}$ arcmin$^{-2}$.
The background map is then the product of this value with the 
exposure map calculated in the same energy band, 0.5--2.0 keV.

\begin{table}
\caption[]{Background estimates in the \rosat PSPC image measured 
in $45^{\circ}$-sectors between 40\arcmin\ and 45\arcmin\ 
from the X-ray emission center.
The number of the Sectors refers to Fig.~\ref{fig:asm}. 
The $\ast$ indicates the regions used to estimate the local 
background. }
\begin{tabular}{lc}
Sector & bkg ($10^{-4}$ cts s$^{-1}$ arcmin$^{-2}$) \\
1 $\ast$ & $4.11 \pm 0.18$  \\
2 $\ast$ & $3.91 \pm 0.17$  \\
3 & $4.39 \pm 0.18$  \\
4 & $5.20 \pm 0.19$  \\
5 & $7.49 \pm 0.39$  \\
6 & $5.04 \pm 0.19$  \\
7 & $4.78 \pm 0.19$  \\
8 $\ast$ & $4.15 \pm 0.18$  \\
\end{tabular}
\end{table}

To characterize the azimuthal deviations in the X-ray surface 
brightness, we have represented the azimuthally averaged profile
with a $\beta$-model (Cavaliere \& Fusco-Femiano 1976): 
\begin{equation}
S (r) = S_0 \left[1+\left(\frac{r}{r_{\rm c}}
\right)^2 \right]^{0.5 -3\beta}.
\end{equation}
We have binned the profile in bins with width of 30 arcsec and used
the local average background. The radial profile is extracted
up to the outer radius, $r_{\rm out} = 1.51$ Mpc, defined as 
the last radial bin where the signal to noise ratio is above 2.
The best-fit results over the radial range 0--1.51 Mpc provide 
a core radius, $r_{\rm c}$, equals to $1.25 (\pm 0.05)$ arcmin, or
$0.097 (\pm 0.004) h_{50}^{-1}$ Mpc, and 
$\beta = 0.473 (\pm 0.004)$ ($\chi^2_{\nu} = 2.0$ with 36 d.o.f.).
The errors are at the 68.3 per cent confidence level ($1 \sigma$)
and are obtained from the distribution of 500 repetitions of the 
fitting process applied to a surface brightness profile randomly
generated from the original one and taking into account the Poisson 
error in each radial bin.
These results are consistent with previous measurements (see, e.g., 
Mohr, Mathiesen \& Evrard 1999, Neumann and Arnaud 1999).
Moreover, we find evidence of elongation of the surface brightness
when the two dimensions distribution is analyzed. 
Fitting the distribution with an elliptical King 
profile, we find an axial ratio of about 0.75 with a position angle of
190 degree (clockwise from 0 degree at West). Using a single core radius,
the two dimensional fit is consistent with the one dimensional radial 
fitting discussed above. 

\begin{figure}
\psfig{figure=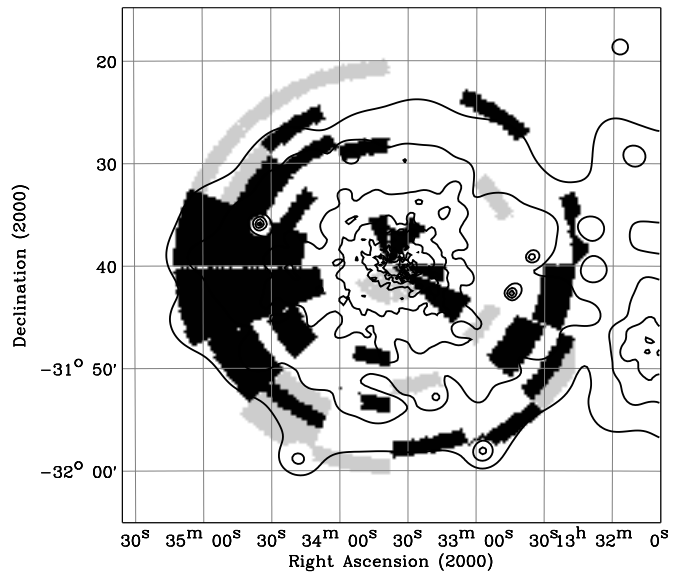,width=.5\textwidth}
\caption[]
{Residuals significant at the 95 per cent confidence level 
once the best-fit $\beta-$model for an azimuthally averaged profile 
is subtracted to the original image. 
The darker regions represent an excess in emission, whereas the lighter
sectors indicate a deficit. 
The overplotted contours are the same of Fig.~\ref{fig:asm}.
Note that 8 point sources are detected and masked.
The source EXO1331.7-3120 (Giommi et al. 1991) appears in the eastern 
sectors.
} \label{fig:res_map} \end{figure}

In Fig.~\ref{fig:res_map}, we investigate the residuals
(significant at the 95 per cent confidence level) obtained when the
original raw image, $I(x,y)$, is compared with a cluster, $S(x,y)$,
simulated by using the best-fit results obtained from the $\beta-$model
added to the local background. 
Both $S(x,y)$ and $I(x,y)$ are in counts pixel$^{-1}$.
The significance of the residuals is assessed in terms of a 
$\chi^2$ distribution (see Press et al. 1992, p.~614):
\begin{equation}
\chi^2 = \sum^{N_x \times N_y}_{j = 1} \frac{[I(x_j,y_j) - 
S(x_j,y_j)]^2} {S(x_j,y_j)},
\end{equation}
where $N_x$ and $N_y$ are the numbers of pixels in the rows and columns in
the selected sectors of 22\fdg5 in annular circular bins of 1.5 arcmin.

The model of the azimuthally averaged surface brightness can account for
about 98 per cent of the total flux from the central 19\arcmin-radius 
region. 
We individuate three main deviations from this circular model:
(i) the core region, which is affected
from the presence of a small cooling flow of $\sim 40 M_{\odot}$ 
year$^{-1}$ (Peres et al. 1998) and from its elliptical shape;
(ii) a slice pointing to South-West, along the axis between A3562 
and SC1329-313; (iii) a region to East, starting 5 arcmin away from 
the X-ray center and extending outward, that produces an excess of 
about 3.3 per cent of the total flux, or a luminosity of about 
$1.8 \times 10^{43}$ erg s$^{-1}$ at the redshift of the cluster. 

To investigate through a $\beta$-model the characteristics 
of the azimuthal profiles, we have extracted the surface 
brightness profiles from eight 45\degr-sectors enclosed between
the ribs and extending up to a radius where the signal 
is above 2 times the uncertainty present in that radial bin.

\begin{figure}
\psfig{figure=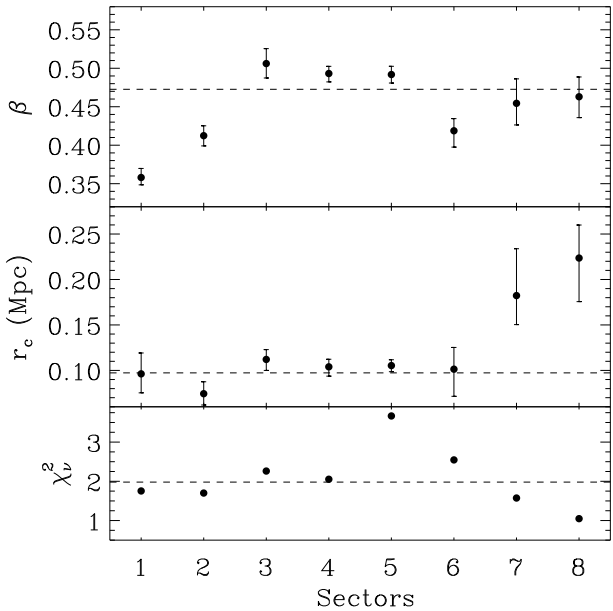,width=.5\textwidth}
\caption[]
{The reduced $\chi^2$, the core radius, $r_{\rm c}$, and the 
$\beta$ value are here plotted versus the 45\degr-sector in which
they are measured. The error bars are significant at 68.3 per cent 
confidence level. The number of the sectors refers to 
Fig.~\ref{fig:asm}.
The dashed lines represent the values measured from the azimuthally
averaged profile.
} \label{fig:sect} \end{figure}

The results of the $\beta$-model fitting are presented in
Fig.~\ref{fig:sect}.
The $\beta$ value for the emission eastward (Sector~1) is 
remarkably smaller ($0.36\pm0.01$) than the value for the azimuthally
averaged profile. 
In Sectors~7 and 8 (located East and South-East),
the values of the core radius are larger than the value obtained
from the azimuthally averaged profile of 0.097 Mpc.
Considering that the parameters of the $\beta$-model correlate
between them, so that higher $\beta$ values corresponds to 
higher $r_{\rm c}$, we conclude that both of these results 
support the evidence for a much flatter distribution of the gas
eastward, probably due to the broadening of the plasma for the merging 
action along that direction.
Finally, we note how poor is the representation with a $\beta-$model
of the cluster emission in Sector~5, which is located in opposition 
to Sector~1 and covers the area in the direction toward the neighboring 
groups.
This region also appears as a deviation at 95 per cent confidence level
from the azimuthally-averaged profile in the map of residuals in
Fig.~\ref{fig:res_map}.

\subsection{Spectral analysis}

We use the REV2 events file, {\tt rp800237n00}, 
processed on January 10th, 1997 from the original dataset.
The main correction applied in the REV2 data that affects the 
spectral analysis is an updated boresighting in the attitude solution
that corrects a mean offset of about 6\arcsec\ observed in the PSPC events
(see \rosat News n.49 \footnote[1]{
available at http://wave.xray.mpe.mpg.de/rosat/mission/rosat\_news}).
The REV2 events file is then corrected for temporal and spatial
gain variations with the {\sc Ftools/Rosat} Perl
script {\it pcpicor 2.2.0} (Snowden et al. 1995, Turner et al. 1995).

We make use of the redistribution
matrix file {\tt pspcb~\_gain2~\_256.rmf} and calculate with 
{\it pcarf 2.1.1} the ancillary response file for each source spectrum.
Another {\sc Ftools/Rosat} Perl script, {\it pcbgdcor 1.0.0}, 
has been used to create 
background spectra with given source region file and 
background region file that allow a proper vignetting correction. 
The original background spectrum is obtained from a circular region
with a radius of 4\arcmin\ at $\sim 45$ arcmin eastward (cfr. Sector~1 
in Fig.\ref{fig:asm}), to avoid any contamination from the 
interacting clumps to the West.
The results of this spectral analysis are discussed below.
We note that these results are consistent with those obtained 
both using background spectra obtained from different regions and 
from the simultaneous fitting of the source and 
background spectra using a model
for the background (see, e.g., Ebeling, Mendes de Oliveira, White 1995)
that combines a power law with photon index 1.97 
to account for the particle background (Snowden et al. 1992), 
a thermal component at 0.124 keV and observed power law with photon 
index 2.12 for the solar scattered photons (Snowden \& Freyberg 1993)
and unresolved AGN (Hasinger 1992), respectively.

The Pulse-Invariant channels are grouped to collect a minimum
of 20 counts in each energy bin to allow the use of the $\chi^2$
analysis.
The spectra are then fitted over the energy range 0.2--2.0 keV to avoid
energy regions with known poor calibration (see Table~3.3 in
Section~3.4 of the \rosat User's Handbook)

\begin{figure}
\psfig{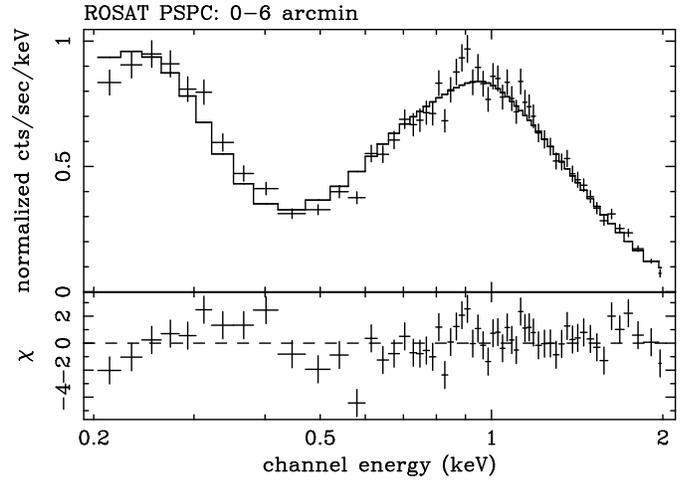}
\caption[] {Best-fit and residuals in $\Delta \chi^2$ of the
PSPC spectral analysis. See Table~3 for best-fit parameters.
} \label{fig:pspc} \end{figure}

The results of the spectral analysis are shown in Table~3.
Due to the low spectral capabilities of the PSPC, we have fixed the metal
abundance to 0.4 times the solar value (Anders \& Grevesse 1989)
as obtained in the \beppo analysis (see next section). 
In the central 6\arcmin-radius region, we measure a temperature
of $3.7^{+0.8}_{-0.6}$ keV (90 per cent confidence level),
a flux of $1.2 \times 10^{-11}$ erg cm$^{-2}$ s$^{-1}$ in
the 0.1--2.4 keV band, with a corresponding un-absorbed luminosity of
$1.5 \times 10^{44}$ erg s$^{-1}$.
In fitting the PSPC spectra, we leave the galactic absorption free to vary
to improve the goodness of the fit (Fig.~\ref{fig:pspc}).
If we fix the absorption to the galactic value of 3.84 $\times 10^{20}$
cm$^{-2}$ (Dickey \& Lockman 1990), 
we measure a temperature of about 2.7 keV but the fit 
is not statistically acceptable ($\chi^2_{\nu} \sim 4.7$ with 174 d.o.f.).

The temperature obtained from \rosat PSPC is in agreement 
with a previous 2--10 keV {\it EXOSAT} observation (Edge \& Stewart 
1991): $kT = 3.8^{+1.0}_{-0.8}$ keV (90 per cent c.l.).
However, as we discuss in subsection~3.1.1, these temperature
estimates disagree (probably for two different reasons) with
measurements obtained from high band-pass X-ray instruments.

\begin{table*}
\caption[]{Best-fit spectral parameters. The errors are at $1 \sigma$
confidence limit. The Galactic value is used when the absorption is 
fixed.  The model used in {\sc Xspec} is {\tt phabs(mekal)}. 
}
\begin{tabular}{cccccc}
data & ring & kT & abundance & $N_{\rm H}$ & $\chi^2_{\nu}$ (d.o.f.) \\
 & arcmin & keV & $Z/Z_{\odot}$ & $10^{20}$ cm$^{-2}$ & \\
 & & & & & \\
PSPC & 0--6 & $3.69^{+0.46}_{-0.36}$ & 0.4 (fix) &              
$2.03^{+0.05}_{-0.05}$ & 1.16 (173) \\
 & & & & & \\
LECS + MECS & 0--8 & $5.13^{+0.21}_{-0.19}$ & $0.39^{+0.05}_{-0.05}$ 
& 3.84 (fix) & 1.09 (210) \\
 & & & & & \\
LECS (0.1--4 keV) & 0--8 & $5.46^{+0.77}_{-0.67}$ & $0.62^{+0.77}_{-0.48}$
& $3.80^{+0.73}_{-0.58}$  & 1.13 (83) \\
LECS (0.2--2 keV) & 0--8 & $>5.42$ (90 per cent c.l.) & $0.4$ (fix)
& $3.26^{+1.03}_{-0.73}$  & 1.02 (53) \\
 & & & & & \\
MECS & 0--8 & $5.10^{+0.23}_{-0.20}$ & $0.39^{+0.05}_{-0.05}$ & 3.84 (fix) 
& 1.09 (124) \\
MECS & 0--2 & $5.26^{+0.50}_{-0.38}$ & $0.50^{+0.10}_{-0.10}$ & 3.84 (fix) 
& 1.21 (69)\\
MECS & 2--4 & $5.74^{+0.43}_{-0.40}$ & $0.36^{+0.09}_{-0.08}$ & 3.84 (fix) 
& 1.17 (78)\\
MECS & 4--6 & $4.97^{+0.50}_{-0.46}$ & $0.37^{+0.11}_{-0.11}$ & 3.84 (fix) 
& 1.16 (60)\\
MECS & 6--8 & $4.60^{+0.76}_{-0.60}$ & $0.30^{+0.20}_{-0.17}$ & 3.84 (fix) 
& 0.91 (60)\\
MECS & 8--12 & $3.22^{+0.55}_{-0.43}$ & $0.12^{+0.14}_{-0.12}$ & 
3.84 (fix) & 0.78 (53)\\
MECS & 12--16 & $3.49^{+0.82}_{-0.65}$ & $<0.16$ & 3.84 (fix) 
& 1.31 (40) \\
MECS & 16--20 & $2.93^{+0.80}_{-0.67}$ & $0.08^{+0.79}_{-0.08}$ 
& 3.84 (fix) & 1.14 (28) \\
\end{tabular}
\end{table*}

\section{\beppo MECS and LECS data}
The cluster A3562 was observed by the \beppo satellite (Boella et
al. 1997a) between 1999 January 31 and February 1.  We discuss here
data from two of the instruments onboard \beppo$\!$: the Medium-Energy
Concentrator Spectrometer (MECS) and the Low-Energy Concentrator
Spectrometer (LECS).  The MECS (Boella et al. 1997b) is presently
composed of two units, working in the 1--10 keV energy range. At
6~keV, the energy resolution is $\sim 8\%$ and the angular resolution
is $\sim$0.7$^{\prime}$ (FWHM).  The LECS (Parmar et al. 1997),
consists of an imaging X-ray detector, working in the 0.1--9 keV
energy range, with 20$\%$ spectral resolution and 0.8$^{\prime}$
(FWHM) angular resolution (both computed at 1 keV).  Standard
reduction procedures and screening criteria have been adopted to
produce linearized and equalized event files.  The MECS (LECS) data
preparation and linearization was performed using the {\sc Saxdas}
({\sc Saxledas}) package under {\sc Ftools} environment.  
The observed count rate for A3562 is 0.174$\pm$0.002 cts s$^{-1}$ for 
the 2 MECS units and 0.133$\pm$0.003 cts s$^{-1}$ for the LECS.
Here we note that the counts rate observed in the high energy 
instrument on board of \beppo$\!$, i.e. the Phoswich Detection 
System (PDS, 15--200 keV), is consistent with zero and presents 
a 90 per cent upper limit of 0.13 cts s$^{-1}$.

\begin{figure}
\psfig{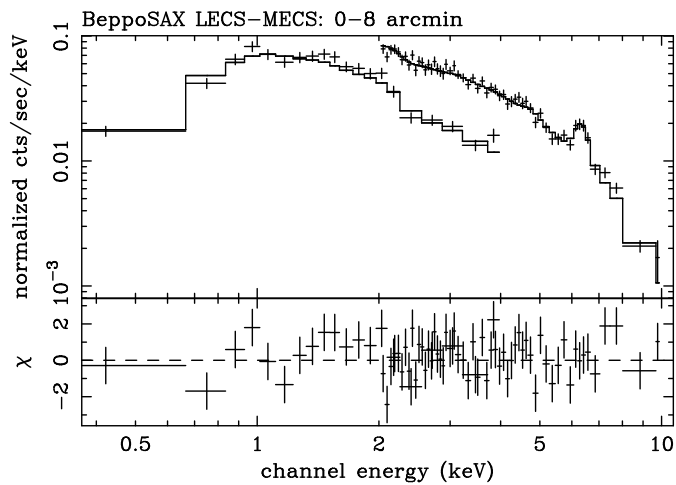}
\caption[] {MECS and LECS spectra and best fitting MEKAL model for 
A3562.  The
MECS and LECS data are extracted from a circular region with a radius
of 8$^{\prime}$ corresponding to $\sim$ 0.62 Mpc.  To improve the
quality of the figure we have rebinned the MECS and LECS spectra in a
number of bins substantially smaller than that used to derive the best
fit reported in the text.
} \label{fig1sax} \end{figure}

\subsection{Broad Band Spectroscopy}
Considering the present status of the calibration of the \beppo instruments
and following the suggestions in Fiore, Guainazzi, Grandi (1999),
we have extracted a MECS spectrum, in the 2--10 keV band, and a LECS
spectrum, in the 0.1--4 keV, both from a circular region of
8$^{\prime}$ radius ($\sim$ 0.62 Mpc), centered on the emission
peak. From the \rosat PSPC radial profile, we estimate that about
70 per cent of the total cluster emission falls within this radius.  The
background subtraction has been performed using spectra extracted from
blank sky event files in the same region of the detectors as the
source.  The counts are then grouped to be a minimum of 25 per bin to
apply properly the $\chi^2$ statistics in the fitting analysis.

The spectra from the two instruments have been fitted simultaneously
with an optically thin thermal emission model (MEKAL code in the 
{\sc Xspec} package), absorbed by a galactic line of sight equivalent 
hydrogen column density, $N_{\rm H}$, of 3.84 $\times 10^{20}$ cm$^{-2}$ 
(Dickey \& Lockman 1990).  
A numerical relative normalization factor among the
two instruments has been added to account for the slight mismatch in
the absolute flux calibration of the MECS and LECS response matrices
employed in this paper (September 1997 release).  The MECS vignetting
is included in the response matrix thanks to the {\sc Effarea} program
(see Molendi et al. 1999 for details).  For the LECS we have used two
redistribution matrices and ancillary response files, the first computed
for an on-axis pointlike source and the second for a source with a
flat brightness profile.  The temperatures and abundances we derive in
the two cases do not differ significantly, as the telescope vignetting
in the 0.1--4.0 keV energy range is not strongly dependent upon energy.
As expected, the relative normalization factor is significantly
different, 0.55$\pm$0.02 in the case of the fit with the point source
effective area, and 0.77$\pm$0.02 in the case of the diffuse source
effective area, where the effects of vignetting are included.

The MEKAL model yields an acceptable fit to the data, $\chi^2 =$ 229.1
for 210 d.o.f. (see Figure~\ref{fig1sax}).
The best fitting values for the temperature and the metal abundance 
are 5.13$^{+0.21}_{-0.19}$ keV and 0.39$\pm$0.05 solar units 
(Anders \& Grevesse 1989), respectively. 
These values are consistent with the results obtained fitting the 
single spectrum of LECS and MECS.

From the best-fit results (cfr. Table~3), we measure an un-absorbed 
flux of $2.0 \times 10^{-11}$ ($2.1 \times 10^{-11}$) 
erg cm$^{-2}$ s$^{-1}$ in the 2--10 (0.1--2.4) keV band with a 
corresponding rest-frame luminosity of $2.1 \times 10^{44}$ ($2.1 
\times 10^{44}$) erg s$^{-1}$. The bolometric luminosity is 
$4.3 \times 10^{44}$ erg s$^{-1}$.

\subsubsection{Comparison with previous estimates}
\begin{figure}
\psfig{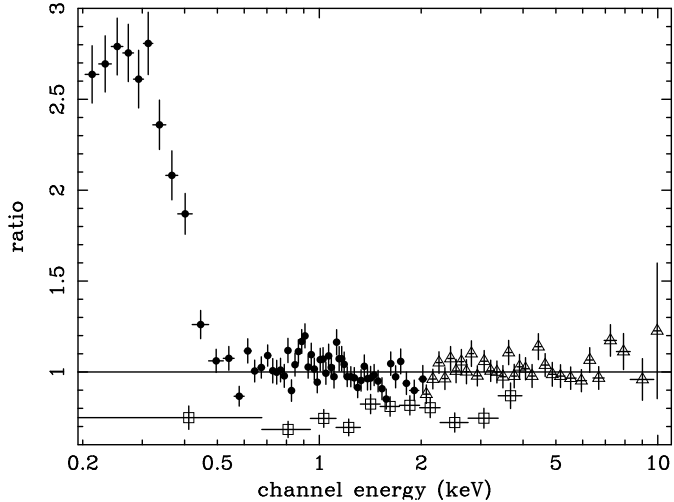}
\caption[] {The LECS ({\it open squares}) and PSPC ({\it filled dots}) 
counts are here divided by the best-fit absorbed MEKAL model with 
fixed galactic absorption obtained from the MECS data ({\it open 
triangles}).  Note that (i) the LECS data are
shifted by a factor $\sim 0.77$ in normalization with respect to MECS
(see text) and (ii) the PSPC counts present a large disagreement
below 0.5 keV.
} \label{fig:ratio} \end{figure}

The estimate of the \beppo broad-band gas temperature is not consistent
with both the \rosat and {\it EXOSAT} estimates by $2.9 \sigma$
and $2.1 \sigma$, respectively (see Section~2.2 and 3.2). 
Once the absorption in front of the thermal model is fixed to the Galactic 
value, the \rosat measurement provides a larger disagreement with the 
\beppo result. In particular, if we extrapolate the MECS results to the \rosat
energy band, we observe large deviations below 0.5 keV
(Fig.~\ref{fig:ratio}).
Therefore, we select the PSPC counts between 0.5 and 2 keV (instead 
of 0.2--2.0 keV as done for the analysis presented in Section~2.2). 
Doing this, the reduced $\chi^2$ value diminishes to 0.97 with 
144 d.o.f. (the improvement is significant to the 87 per cent confidence
level, using the F-test), but still the best-fit gas temperature 
remains low ($kT = 3.1 \pm 0.3$ keV, 1 $\sigma$; note that 
if we fix the parameters of the spectral fit to the best values
obtained from the \beppo spectra --$kT=5.3$ keV, $Z=0.4 Z_{\odot}$ 
and galactic absorption-- the $\chi^2_{\nu}$ is 1.14 and can not be excluded
at 0.05 significance level given the $\chi^2$ probability distribution).
Finally, if we fit the LECS spectrum between 0.2 and 2 keV, we obtain
a reasonable good agreement with both the Galactic absorption and
the temperature estimate as measured from MECS data only (cf. Table~3).

On the other hand, using the Japanese-US satellite {\it ASCA} 
that operates in a similar energy band as the \beppo MECS instrument, 
Hanami et al. (1999) and White (2000) observe a gas temperature in A3562 of 
$5.3 \pm 0.2$ and $5.2 \pm 0.2$ keV, respectively, 
with a metallicity of about 0.3 times the solar value 
and fixing the absorption to the galactic value. 
Moreover, White (2000) finds a statistical evidence (90 per cent confidence
level) that a cooling flow component is required from the {\it ASCA}
GIS spectra. Including this component, the ambient gas temperature 
raises to $7.0^{+1.8}_{-1.0}$ keV. 

About the {\it EXOSAT} measurement, it is worth to note that
the spectrum was obtained from the Medium-Energy array of proportional
counters that has no spatial resolution and a quite large field-of-view
of 45\arcmin$\times$45\arcmin. We have verified that the observation 
has been centered between A3562 and SC1329-313, at (RA, Dec) =
13$^{\rm h}$ 32$^{\rm m}$ 14\fs4, --31\degr 41\arcmin 22\arcsec, 
and covers only 65 per cent of the central circular region of the
PSPC centered on the cluster. 
Therefore, a significant contamination from the cooler group of galaxies
could have affected the measurement of the cluster temperature.
 
In principle, a possible explanation of this soft excess can be the fact
that A3562 is located not far from the galactic center,
$(l, b) \sim$ (313\fdg3, 30\fdg4), where the \rosat soft band maps
of Snowden et al. (1997) show a strong contribution from 
the Galactic bulge and the SNR/stellar wind bubble Radio Loop I.
However, if this foreground contamination is homogeneous, diffuse
and extended over the field of view of the exposition, 
the use of a local background should correct for this excess.
We have tried to fit the \rosat spectra with a two temperatures
model, but we have not obtained any reduction of the $\chi^2$.
Furthermore, we do not observe any disagreement between the best-fit
MECS results and the extrapolation to the LECS energy band (note that
the {\it squares} in Fig.~\ref{fig:ratio} are consistent with a constant
value).

Finally, we note that the same trend of a lower value of the plasma
temperature as measured from \rosat PSPC than the estimate
obtained from spectra collected by instruments with wide 
X-ray energy band (e.g. MECS, {\it ASCA} GIS and SIS) has been also 
observed in A3558, the most massive cluster in the Shapley region:
from the PSPC data, Bardelli et al. (1996) measure 3.3 keV that has 
to be compared with temperatures larger than 5 keV obtained from 
{\it ASCA} [$5.5 \pm 0.4$ keV in Markevitch et al. (1998); 
$6.0 \pm 0.2$ keV in Hanami et al. (1999);
$5.5 \pm 0.1$ keV in White (2000)].

Thus, given also the reasonable agreement between {\it ASCA} and 
\beppo measurements, it seems that a calibration problem affects the 
soft \rosat band, i.e. at energies between 0.2 and 0.5 keV where
large positive residuals are still observed also for calibration sources 
due to temporal variation of the PSPC gain (see section~B.2 of the 
\rosat User's Handbook, Prieto et al.  1994, Markevitch \& Vikhlinin 1997).
Here, we note that there are other pieces of evidence that support 
this conclusion from an observational point of view: 
Allen \& Fabian (1997) note that \rosat PSPC spectra
provide generally lower excess absorption when compared with {\it ASCA} 
results; Iwasawa, Fabian \& Nandra (1999) measure a significant 
disagreement between PSPC and {\it ASCA} 
in the soft spectrum of a simultaneous
observation of NGC~5548 (cfr. their Fig.~7 with our Fig.~\ref{fig:ratio});
Mineo et al. (2000) find same evidence of PSPC soft excess when compared
to LECS data in a sample of PG quasars.

\subsection{Spatially Resolved Spectral Analysis}

The distribution of the gas temperature within the clusters
provides a powerful tool to investigate the dynamical status
of the intracluster plasma and to constrain the gravitational mass
in hydrostatic equilibrium with the gas itself.

A combined good spatial and spectral sensibility is necessary to 
resolve the gas temperature distribution. The \rosat PSPC has 
a good spatial resolution but is spectrally limited to energy
below 2.4 keV. On the other hand, {\it ASCA} is sensitive to 
photons with energies up to 10 keV typical for a hot cluster, 
but its Point-Spread-Function (PSF) has a large ($\sim$3 arcmin) 
half-power diameter that depends strongly upon the energy
(Serlemitsos et al. 1995).
A better combination of the properties required is provided from
the MECS instrument onboard \beppo$\!$. 
The PSF of the MECS has a half-power diameter of 2.5 arcmin and is found
to vary only weakly with energy (D'Acri, De Grandi \& Molendi 1998).
Nonetheless, we have taken into account the small spectral distortions 
introduced by the PSF using appropriated effective area files
produced with the {\sc Effarea} program for each annulus and
sector in which a source spectrum has been accumulated.

\begin{figure}
\psfig{figure=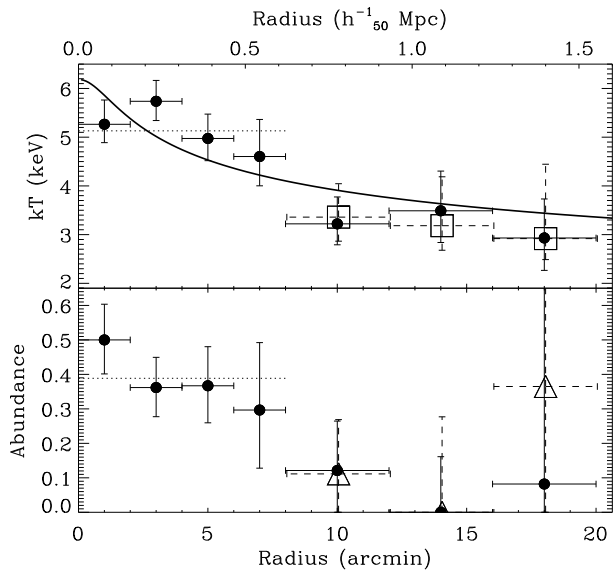,width=.5\textwidth}
\caption{
{\bf Top Panel}: projected radial temperature profile.  The filled
circles and the open squares indicate respectively the measurements
obtained by fitting the continuum spectrum including and excluding the
point source at $\sim 13^\prime$ from the cluster peak.
The dotted line between 0$^\prime$ and 8$^\prime$ shows the best
spectral fit values from LECS and MECS counts collected over this region.
The solid line represents the best polytropic fit to the temperature
profile. 
{\bf Bottom Panel}: projected radial abundance profile. The filled
circles and the open triangles indicate respectively the measurements
obtained by fitting the continuum spectrum including and excluding the
point source.
} \label{fig2sax} \end{figure}

\subsubsection{Radial Profiles}
We have accumulated spectra from seven annular regions centered on the
X-ray emission peak (see Fig.~\ref{fig2sax} and Table~3). 
The background subtraction has been performed using spectra 
extracted from blank sky event files in the
same region of the detector as the source.  A correction for the
absorption caused by the strongback supporting the detector window has
been applied for the 8$^{\prime}$-12$^{\prime}$ annulus, where the
annular part of the strongback is contained. For the
4$^{\prime}$-6$^{\prime}$, 12$^{\prime}$-16$^{\prime}$ and the
16$^{\prime}$-20$^{\prime}$ annuli, where the strongback covers only a
small fraction of the available area, we have chosen to exclude the
regions shadowed by the strongback.  For the 4 innermost annuli the
energy range considered for spectral fitting is 2--10 keV. 
The 8$^{\prime}$-12$^{\prime}$ annulus is the more sensitive to 
the strongback correction. This correction is less reliable at low 
energies: if we use 2 and 3 keV as lower energy cutoff,
the best-fit results provide a temperature of about 5 and 3.4 keV,
respectively. In the present analysis, we have chosen to accumulate
photons with energies between 2.7 and 10 keV and to flag this bin 
for a systematic uncertainties that can be up to twice the
statistical error.
Finally, for the 2 outermost annuli, the fit is restricted to the 
2--8 keV energy range. The softer energy range considered for the 
outer annuli limits the spectral distortions, which could be caused 
by an incorrect background subtraction. 

We have fitted each spectrum with an absorbed MEKAL model. 
For the spectrum
from the 0$^{\prime}$-2$^{\prime}$ region we have also performed a fit
with a one temperature thermal emission component plus a cooling flow
component (MKCFLOW code in the {\sc Xspec} package). The parameters of this
component have all been fixed, as the energy range we use for spectral
fitting (2--10 keV) is not particularly sensitive to the cooling flow.
More specifically, the minimum temperature is fixed at 0.1 keV, the
maximum temperature, and the metal abundance is set to be equal to
the temperature and the metal abundance of the MEKAL component, the
mass deposition rate, $\dot M$ is fixed to 37 M$_\odot$ yr$^{-1}$, 
(from the deprojection analysis of the 0.4--2.0 keV \rosat PSPC
image in Peres et al. 1998).  The derived values for the
ambient temperature and the metal abundance are indistinguishable from
those derived for the single temperature analysis.

In Fig.~\ref{fig2sax}, we show the temperature and abundance 
profiles (filled circles) obtained from the spectral fits. 

By fitting the temperature and abundance profiles with a constant 
we derive the following average values: $4.67\pm$0.22 keV and 
0.33$\pm$0.05 solar units.  
A constant provides an acceptable fit to the abundance profile,
$\chi^2 =$ 9.4 for 6 d.o.f. (probability = 0.15).
On the other hand, a constant does not provide an acceptable fit 
to the temperature profile (Table~4).  Using
the $\chi^2$ statistics we find $\chi^2 =$ 21.7 for 6 d.o.f.,
corresponding to a probability of 0.001 for the observed distribution
to be drawn from a constant parent distribution.  
We have tried to fit a more physical functional form, obtained 
from the assumption of a polytropic relation between the gas density,
described by a $\beta-$model, and temperature:
\begin{equation}
T(r) = T_0 \left[1+\left(\frac{r}{r_{\rm c}}\right)^2 \right]^{-1.5\beta
(\gamma -1)}.
\end{equation}
Using the best-fit values for the core radius, $r_{\rm c}$, and $\beta$
(see Section~2.1), we measure $kT_0 = 6.19^{+0.53}_{-0.59}$ keV
and $\gamma = 1.16 \pm 0.03$ ($\chi^2 =$8.6 with 5 d.o.f.), i.e.
a significant 4.8 $\sigma$ deviation from an isothermal profile.
(The error bars come from the distribution of the parameter values
after 500 repetitions of the polytropic fit, each time performed 
with a new set of [$r_{\rm c}, \beta$]; random values of the 
temperature and surface brightness profiles are considered for the fit
with a polytropic and $\beta$ model, respectively, 
according to the error estimates at each radial bin of the two profiles).
We note that, from the deconvolution of the {\it ASCA} data, White (2000)
measures a flat profile up to $\sim$17 arcmin. However, the last bin 
covers the radial range between 6.4 and 16.7 arcmin and, thus, is not able 
to resolve any gradient.

\begin{figure}
\psfig{figure=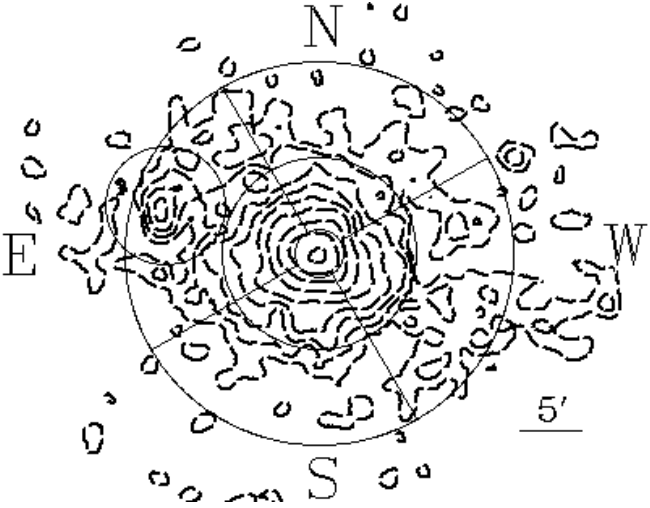,width=.5\textwidth,angle=0}
\caption{\beppo MECS image of A3562. Logarithmic contour levels are
indicated by the dashed lines. The solid lines show how the cluster
has been divided to obtain temperature and abundance maps. The 
circle in the eastern sector indicates the region excluded around
EXO1331.7-3120 when the spectra are accumulated
without this point source.
} \label{fig3sax} \end{figure}

The \rosat PSPC and MECS image (see Fig.~\ref{fig3sax}) of A3562 contains 
an unrelated strong point source (EXO1331.7-3120, serendipitously 
detected with {\it EXOSAT}, Giommi et al. 1991), 
located at $\sim 13$ arcmin from the center of the cluster. 
We have accumulated the
spectra excluding a region of 5$^{\prime}$ around this point source
and rederived the temperature and metal abundance for the 3 outer
annuli. Our results are over-plotted in Fig.~\ref{fig2sax} both for the
temperature (open squares in the top panel) and abundance (open
triangles in the bottom panel) measurements. Clearly, the excision
of the point source from the spectra, does not modify in any substantial
way the measured temperatures and abundances.

Finally, if we fit the spectra of the last two bins over a harder energy 
range (2--10 keV instead of the adopted 2--8 keV), we obtain that
the spectral distorsion due to an incorrect background subtraction
decreases the best-fit central values in the two bins by 25 and 50 per cent,
respectively, with a relative uncertainties of about 50 per cent.
 
\begin{figure}
\psfig{figure=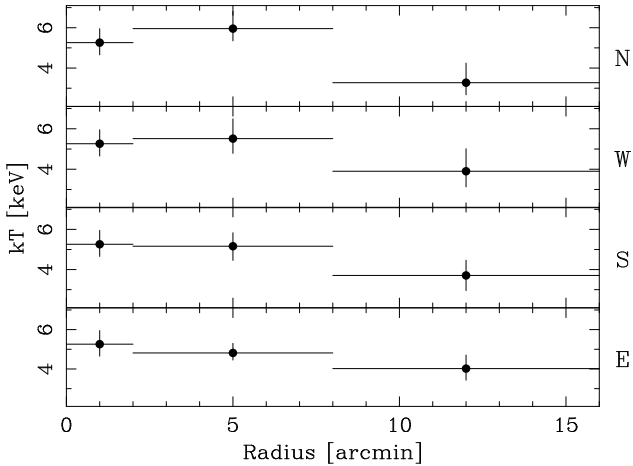,width=.5\textwidth,angle=-90}
\caption{Radial temperature profiles for the north
sector (first panel), the west sector (second panel), the
south sector (third panel) and the east sector (forth panel).
The temperature for the inner bin is derived from the entire
circle, rather than from each sector. 
In Table~4, we quote the results of the best-fit of these profiles  
with a constant.  }
\label{fig4sax} \end{figure}

\subsubsection{Analysis by sectors}
We have divided A3562 into 4 sectors: north, west, south and east.
Each sector has been divided into 2 annuli with bounding radii
2$^{\prime}$-8$^{\prime}$ and 8$^{\prime}$-16$^{\prime}$. 
In Fig.~\ref{fig3sax},
we show the MECS image with the sectors overlaid.  The background
subtraction has been performed using spectra extracted from blank sky
event files in the same region of the detector as the source.  A
correction for the absorption caused by the strongback supporting the
detector window has been applied for the sectors belonging to the
8$^{\prime}$-16$^{\prime}$ annulus. We have adopted the 2--10 keV
energy range for spectra from all annuli.  In the case of the spectra
from the outermost annulus, 8$^{\prime}$-16$^{\prime}$, we have
verified that reducing the energy range to the 2--8 keV band does not
alter in any significant way the best fitting values for the
temperature and metal abundance.  We have fitted each spectrum with a
MEKAL model absorbed by the galactic line of sight equivalent hydrogen
column density.

\begin{figure}
\psfig{figure=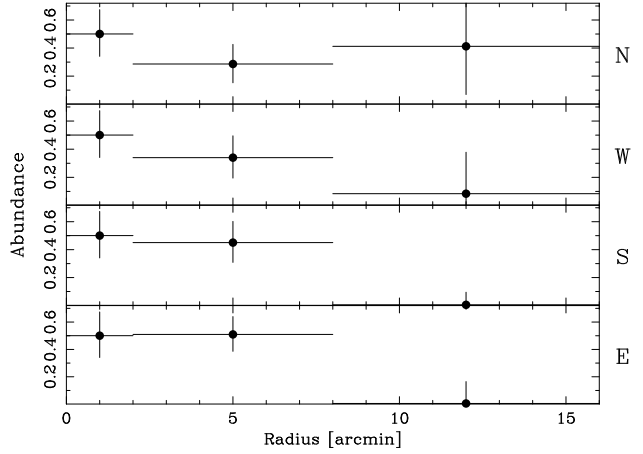,width=.5\textwidth,angle=-90}
\caption{Radial abundance profiles for the north sector (first panel),
the west sector (second panel), the south sector (third panel) and the
east sector (forth panel).  The abundance for the inner bin is
derived from the entire circle, rather than from each sector.
In Table~4, we quote the results of the best-fit of these profiles
with a constant.  }
\label{fig5sax}
\end{figure}

\begin{table}
\caption[]{Best-fit constant values for gas temperature and metal abundance 
profiles observed in the four sectors in which we mapped the \beppo 
observation and in azimuthally averaged annuli ('All').
}
\begin{tabular}{lcccccccc}
Sector && $kT$ (keV) & $\chi^2_{\nu}$ && $Z/Z_{\odot}$ & $\chi^2_{\nu}$ 
&& d.o.f. \\
\\
North && 5.0$\pm$0.5 & 2.3 && 0.37$\pm$0.11 & 0.5 && 2 \\
West  && 5.1$\pm$0.5 & 0.7 && 0.37$\pm$0.11 & 0.8 && 2 \\
South && 4.8$\pm$0.4 & 1.4 && 0.11$\pm$0.05 & 5.2 && 2 \\
East  && 4.7$\pm$0.4 & 0.9 && 0.36$\pm$0.09 & 3.5 && 2 \\
\\
All   && 4.7$\pm$0.2 & 3.6 && 0.33$\pm$0.05 & 1.6 && 6 \\
\end{tabular}
\end{table}

In Fig.~\ref{fig4sax}, we show the temperature profiles obtained from the
spectral fits for each of the 4 sectors.  Note that in all the
profiles we have included the temperature measure obtained for the
central circular region with radius 2$^{\prime}$.  
All four sectors show a decrease of the temperature at large radii,
consistent with the trend shown in Fig.~\ref{fig2sax} for the entire
cluster. Moreover, when fitted with a constant temperature, 
the resulting best fit temperature for each sector is consistent 
at the 90 per cent confidence level with the average
temperature for A3562 derived in the previous section.

In Fig.~\ref{fig5sax}, 
we show the abundance profiles for each of the 4 sectors.
In all profiles we have included the abundance measure obtained for
the central circular region with bounding radius 2$^{\prime}$.
We have fitted a constant abundance to these profiles (see Table~4).
While in the south and east sectors we have clear evidence of an 
abundance decline with increasing radius, the modest statistics 
of the 8-16 arcmin annulus in the north and west prevent us 
from drawing any conclusion in these sectors.

\section{Constraints on the masses}

\begin{figure}
\psfig{figure=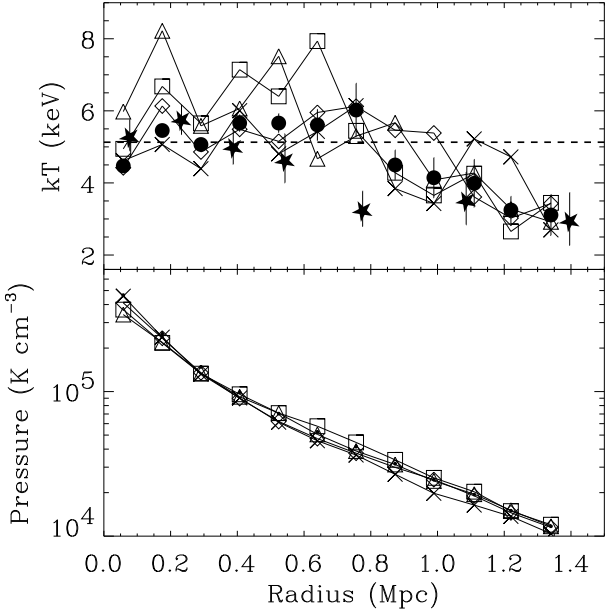,width=.5\textwidth}
\caption[]{This plot shows the dependence upon the distance
from the center of the X-ray emission of the gas temperature
and thermal pressure as obtained from the deprojection of the
azimuthally averaged profile. 
To underline the deviations from the latter results, 
we use the same input parameters 
to deproject the four 90\degr--width azimuthal regions:
North--{\it cross}, East--{\it square}, South--{\it triangle},
West--{\it diamond}. 
The {\it stars} represent the temperature profile obtained from
the \beppo spectra, and the {\it dots} the deprojected temperature
profile.
Note that the temperature value in the 8$^{\prime}$-12$^{\prime}$ annulus 
is affected by a systematic error twice as large as the plotted 
statistical uncertainty (see comments in Section~3.2.1).
} \label{fig:deproj}
\end{figure}

We have directly deprojected the surface brightness from the 
azimuthally averaged profile and four azimuthal
sectors to look for variations in the gas temperature and pressure
(see Fig.~\ref{fig:deproj}) and to constrain the distribution of the
gas and total mass. 

The deprojection technique (Fabian et al. 1981, White et al. 1997)
has been applied to the surface brightness profiles extracted 
from exposure--corrected image in the 0.5--2.0 keV band. 
Given the assumption that the observed projected cluster emission is due
to the X-ray emitting gas in spherically symmetric shells,
the count emissivity in each radial volume shell can be calculated
and compared with the predicted counts from a
thermal emission of an optically thin gas [described by a 
MEKAL model, based on the model calculations of Mewe and Kaastra
(Kaastra 1992) with Fe L calculations by Liedahl et al. (1995)], 
absorbed by intervening matter ($N_{\rm H} = 3.84 \times 10^{20}$ 
cm$^{-2}$) and convolved with the response of the detector. 
For such an emission model, the calculated flux is proportional to the
electron density, $n_{\rm e}$, and the intracluster medium
temperature, $T_{\rm gas}$, according to the relation: $n_{\rm e}^2
f(T_{\rm gas})$ (where $f(T_{\rm gas}) \stackrel{\propto}{\sim} T_{\rm
gas}^{0.5}$, for $T_{\rm gas} \ga 3\times10^7$ K). 
Applying the perfect gas law and the equation of hydrostatic equilibrium
to the ICM and assuming a form for the gravitational potential, 
the temperature and density profiles are obtained once, e.g.,
the pressure in the outermost bin is fixed.

The gravitational potential functional form is defined by the 
Navarro, Frenk \& White dark matter profile [\(\rho_{\rm DM} \propto
x^{-1}(1+x)^{-2} \), where $x=r/r_{\rm s}$; Navarro et al. 1995] 
parametrized with the velocity dispersion $\sigma \sim GM/(2r)$ 
and the scale radius, $r_{\rm s}$.

The criterion in obtaining a reasonable deprojection is to match
an assumed or observed gas temperature profile.
We have looked for a good representation of the temperature profile
obtained from our analysis of the \beppo data.
The gas temperature profile is well reproduced from the 
deprojection of the azimuthally averaged profile by choosing
a $\sigma = 800$ km s$^{-1}$ (in good agreement with the 
optical determination of $913^{+189}_{-96}$ km s$^{-1}$, Bardelli 
et al. 1998a), $r_{\rm s} = 0.40$ Mpc and a pressure value at the
outer radius $r_{\rm out} = 1.51$ Mpc of $0.8 \times 10^4$ 
Kelvin cm$^{-3}$.

The results of the deprojection provide a mass deposition rate
of about 33 $M_{\odot}$ year$^{-1}$, with a 90th percentile
upper limit of 70 $M_{\odot}$ year$^{-1}$, which is consistent
with the result in Peres et al. (1998).
 
To make a qualitative comparison among the deprojected profiles
of the four azimuthal sectors, we have fixed the input parameters
to the values above. In Fig.~\ref{fig:deproj}, 
the eastern ({\it squares}) and southern ({\it triangles}) temperature
profiles show a steep behaviour in the core that imply the use of
a larger scale dark matter radius, i.e. a broader distribution of the 
gas, to reproduce the observed profiles in Fig.~\ref{fig4sax}.
(Note that the 8$^{\prime}$-12$^{\prime}$ annulus is the more sensitive to
the strongback correction; cf. second paragraph in Section~3.2.1).

At the penultimate radial bin, $r=1.40 h_{50}^{-1}$ Mpc, we measure
a gas mass, $M_{\rm gas}$, of $9.0 (\pm 0.2) \times 10^{13} h_{50}^{-5/2}
M_{\odot}$,
a total mass, $M_{\rm tot}$, of $4.7 \times 10^{14} h_{50}^{-1} M_{\odot}$,
and a gas fraction, $f_{\rm gas} = M_{\rm gas} / M_{\rm tot}$
of $0.19 (\pm 0.01) h_{50}^{-3/2}$.

Using the results from the analysis of the surface brightness profile
up to 1.51 Mpc done with the $\beta-$model, we can investigate
the properties of A3562 at the representative radius $r_{200}$,
where the mean overdensity in the cluster is 200 times the 
critical density and the virial equilibrium within the cluster
is reached. 
In detail, the gas mass is given by the integration over the
cluster spherical volume of the gas density represented
from the $\beta-$model, the total mass is obtained
from the hydrostatic equilibrium between the cluster potential
and the plasma (see, e.g., Henriksen \& Mushotzky 1986):
\begin{eqnarray}
M_{\rm tot} (r) & = & -\frac{kT(r) \ r}{G \mu m_{\rm p}} \left(
\frac{\partial \ln \rho}{\partial \ln r} + \frac{\partial \ln kT}{\partial
\ln r}\right)  \nonumber \\
 & = & \frac{3 \ \beta \gamma \ kT_0 \ r_{\rm c}}{G \mu m_{\rm p}}
\frac{x^3}{ (1+x^2)^{B} } \nonumber \\
 & = & \frac{1.13 \times 10^{14}}{(\mu/0.59)} \ \beta \ \gamma \ kT_0 
\ r_{\rm c} \frac{x^3}{ (1+x^2)^{B} } \ M_{\odot},
\label{app:mtot}
\end{eqnarray}
where $x = r/r_{\rm c}$, $B = 1.5 \beta (\gamma -1) +1$, $kT_0$ is the central
temperature in keV, $r_{\rm c}$ the core radius in $h_{50}^{-1}$ Mpc,
$\mu$ is the mean molecular weight in a.m.u.
and the numerical values include the gravitational constant $G$,
the mass of the proton $m_{\rm p}$ and all the unit conversions.
Finally, given this mass profile, $r_{200}$ is given by
a simple analytic formula (Ettori 2000).

Under the isothermal assumption (i.e., $\gamma =1$ and $kT(r) = kT_0 =
5.13 \pm 0.21$ keV; see Section~3.1), $r_{200} = 2.02 \pm 0.01$ Mpc,
$M_{\rm gas} = 1.3 (\pm 0.1) \times 10^{14} h_{50}^{-5/2} M_{\odot}$, 
$M_{\rm tot} = 5.5 (\pm 0.2) \times 10^{14} h_{50}^{-1} M_{\odot}$ 
and the gas fraction, $f_{\rm gas}$, is $0.24 (\pm 0.01) h_{50}^{-3/2}$.
Once a polytropic temperature profile is considered
(see best-fit results in Section~3.2.1) and a proper correction
is applied to the projected central temperature ($kT_0$ changes 
from the best-fit value of 6.19 keV to the deprojected value of 
6.63 keV; see Markevitch et al.  1999), $r_{200}$
decreases to 1.79 ($\pm$0.05) Mpc, the total mass at that radius
is $3.9 (\pm 0.4) \times 10^{14} M_{\odot}$,
the gas mass is $1.1 \times 10^{14} M_{\odot}$, and the gas mass 
fraction is $0.28 (\pm 0.03) h_{50}^{-3/2}$.

It is worth to note that these values of the gas and total mass, 
and the relative gas mass fraction, are extrapolations
of the observed masses profiles and are consistent at the 90 per
cent confidence level with
the results of the deprojection analysis over the radial range
covered from the cluster emission.


These estimates of the total mass are generally lower than
the values quoted in Quintana et al. (1995), where 
different estimators of the virial mass applied to the galaxy counts
within 2 Mpc
provide values in the range $[7.7, 17.3] \times 10^{14} M_{\odot}$.
Using the galaxy density distribution and the isothermal assumption, 
Bardelli et al. (1998a) measure a total mass of $6.8 \times 10^{14} 
M_{\odot}$ at $r = 1.5$ Mpc.
This disagreement can be due to the difficulty in disentangling
the cluster properties only through optical analysis for the 
presence of mergers and contaminations that affect the 
estimate of the optical velocity dispersion and the definitions
of the galaxy members. 

\section{DISCUSSION AND CONCLUSIONS}

From the present analysis of the \rosat PSPC and \beppo data,
we conclude that A3562 is a moderate rich cluster with a X-ray
bolometric luminosity of $4.3 \times 10^{44} h_{50}^{-2}$ 
erg s$^{-1}$ in a not-relaxed status, with 
evidence of merging and/or interaction along the West-East axis.
The broader gas emission along that axis is consistent with the
picture described by N-body and hydrodynamical simulations, 
in which the clumps, originally spherical, merge in a cigar-like
shape structure that propagates the shock front in the
outskirts and relaxes (see, e.g., Fig.~3 in Takizawa 1999).
This scenario is complementary and in support of the spectral analysis
of the {\it ASCA} spectra presented by Hanami et al. (1999).

The complex morphology of the cluster is also underlined from
the comparison between the two $\beta$ values, one obtained 
from the spatial analysis, $\beta_{\rm imag}$, and the other
from the spectral and optical data, $\beta_{\rm spec} 
= \frac{\mu m_{\rm p} \sigma_{\rm opt}^2}{kT}$.
Representing the estimate of the same quantity, i.e. 
the ratio of the energies per unit mass stored in the galaxies
and in the gas (Cavaliere \& Fusco-Femiano 1976), these values
should be equal at least in relaxed systems. 
However, deviations in the average values of $\beta_{\rm imag}$ and
$\beta_{\rm spec}$ have been observed, rising the so-called
$\beta-${\it problem} (Mushotzky 1984, Edge \& Stewart 1991, 
Bahcall \& Lubin 1994).
This disagreement is observed more significantly in systems that 
present mergers and complex morphology. 
The case of A3562 confirms this tendency:
given an optical velocity dispersion from Bardelli et al. (1998a)
and the X-ray spectral temperature $kT = 5.13^{+0.21}_{-0.19}$ 
keV obtained
from the \beppo spectra, we measure $\beta_{\rm spec} =
1.00^{+0.42}_{-0.21}$ that is not consistent at the 2.5 $\sigma$
level with $\beta_{\rm imag} = 0.473 \pm 0.004$ obtained from 
the modeling of the surface brightness profile with 
a spherically-symmetric and isothermal X-ray emission.

From the analysis of the \beppo data, we observe a gas 
temperature that is consistent with {\it ASCA} measurements
and significantly higher than the estimate obtained 
from \rosat and {\it EXOSAT}. We suspect that 
a calibration problem affects the measure done with \rosat PSPC
and a misdirect pointing that done with {\it EXOSAT}.
Here we note that also the main cluster in the Shapley core,
i.e. A3558, presents a temperature of the intracluster plasma
lower than the {\it ASCA} estimate (Markevitch \& Vikhlinin
1997, Hanami et al. 1999, White 2000) when it is measured with 
\rosat PSPC (Bardelli et al. 1996).

We observe evidence of a gradient in the temperature and
abundance profiles. The presence of the small cooling flow in the core of
A3562 and the steep decrease of the metallicity in the southern and
eastern sectors may suggest that the cluster is undergoing a successive
merger with a group with a luminosity of about $10^{43}$ erg s$^{-1}$
and poor metallicity in the East, after a time in which the relaxation 
allowed the enhancement of the gas density in the core.
It is worth to note that Bardelli et al. (1998b) show the evidence
of two merging groups in the same directions of the observed
X-ray excess: the clump 'T598' has a peak density at 
17.7 arcmin to East of the X-ray center, while 'T599' 
is separated by 28.9 arcmin from the X-ray center to the South-East.
From the isodensity of the bi-dimensional distribution (Fig.~9 of 
Bardelli et al 1998b) and the redshift distribution (Fig.~2 of Bardelli 
et al 1998a) of galaxies belonging to these groups, they appear 
quite dispersed, probably due to strong tidal interaction with the
more massive A3562.

From a recent survey performed at 20 cm in the region of the A3558
chain (Venturi et al. 2000), it results that A3562 hosts in its
neighborhood nine of the 28 radio sources of the cluster complex.  
Seven of these objects are located in the eastern edge, where the
dynamical analysis found the existence of the two groups, T598
and T599, cited above.
These two groups seem to connect each other near
the position of the X-ray excess we detected at a distance of $\sim
15$ arcmin from the A3562 center.
Two (J1333-3141 and J1335-3153) of the A3562 radio sources 
present extended morphology and it was possible to estimate their
physical properties. 
In particular, the equipartition pressure of J1333-3141
is $P_{r}=0.4 \times 10^{-12} h_{50}^{4/7}$ dyne cm$^{-2}$
(or $2.9 \times 10^3$ K cm$^{-3}$):
if the source is at the projected distance of $101 h_{50}^{-1}$ kpc,
the pressure of the hot gas $P_{hg}$ as estimated from the
deprojection will be a factor 100 higher than $P_{r}$. This is not a
strange result, since it is known that the radio emission could be
overpressured by the diffuse hot gas up to a factor 10-100
(Feretti, Perola \& Fanti 1992).
However, if we impose $P_{hg}=P_{r}$, we derive a physical distance 
of $\sim 0.9$ Mpc of the radio source from the cluster
center, with an inclination of the tail of $\sim$ 83\degr
with respect to the plane of the sky.
More interesting is the source J1335-3153, whose pressure is 
$P_{r}=0.5 \times 10^{-12} h_{50}^{4/7}$ dyne cm$^{-2}$; its
projected distance from the cluster center is $2.3\ h_{50}^{-1}$ Mpc.
At this distance, the gas of A3562 is not able to confine the
extended radio emission, which on the contrary could be in
interaction with the more diffuse component seen in the X-ray map, 
if it extends to the radio source.
A more detailed analysis, comprising also the age of the relativistic
electrons derived from the spectra, is underway.

A \beppo observation of each of the two groups of galaxies present in
the Shapley core, SC1327-312 and SC1329-313, is planned by the end of
the present AO (March 2000). These observations will allow us to
describe in detail the ongoing merging thanks to the capability of
\beppo to map the temperature distribution as shown in the present
work.

\section*{ACKNOWLEDGEMENTS} 
This research has made use of linearized event files produced at the 
\beppo Science Data Center and data obtained through the
High Energy Astrophysics Science Archive Research Center Online 
Service, provided by the NASA-Goddard Space Flight Centre.
SE acknowledges the support of the Royal Society. 
The anonymous referee is thanked for comments that improved the 
presentation of this work.

\end{document}